\def\editmode{0}    
\def\singlenarrowcol{0}

\def\bibfilenames{bibman_refs}

\def\spsformat{0}
\if\singlenarrowcol0
\if\spsformat1

\documentclass{article}
\usepackage{spconf}

\else
\documentclass[10pt,conference]{IEEEtran}

\IEEEoverridecommandlockouts

\fi
\else

\documentclass[onecolumn]{IEEEtran}
    \usepackage[pass]{geometry}
    \newgeometry{textwidth=252pt, textheight=672pt}

\fi

\usepackage{include_v7}

\usepackage{notation}

\newcommand{\cbr}[1]{} 

\usepackage{hyperref}
\usepackage{balance}

\usepackage{tikz}
\usetikzlibrary{arrows}
\usetikzlibrary{fit}

\def\journal{0}

\usepackage{environ}

\if\journal0    
    \NewEnviron{journalonly}{%
    }
    \NewEnviron{conferenceonly}{%
    \BODY
    }
\else
    \if\journal1
        \NewEnviron{journalonly}{%
        \BODY
        }   
        \NewEnviron{conferenceonly}{%
        }  
    \else
        \if\journal2
        \newenvironment{journalonly}{\color{darkyellow}}{}
        
        \fi
    \fi
\fi

\allowdisplaybreaks

\begin{document}

\title{Radio Maps for Beam Alignment in mmWave Communications with Location Uncertainty\thanks{
        This research has been funded in part by the Research Council of Norway under IKTPLUSS grant 311994 and by MCIN/AEI/10.13039/501100011033/FEDER ``A way of making Europe” under project MAYTE (PID2022-136512OB-C22).

    }
}

\author{
\IEEEauthorblockN{Tien Ngoc Ha, Daniel Romero}
\IEEEauthorblockA{\textit{
    Department of ICT},
\textit{University of Agder}\\
Grimstad, Norway \\
{\{tien.n.ha, daniel.romero\}@uia.no}}
\and
\IEEEauthorblockN{Roberto López-Valcarce}
\IEEEauthorblockA{\textit{
    atlanTTic Research Center},
\textit{University of Vigo}\\
Vigo, Spain \\
{valcarce@gts.uvigo.es}}
}
\maketitle

\begin{abstract}
    Next generation communication systems require  accurate beam alignment to
    counteract the impairments that characterize propagation in high-frequency
    bands. The overhead of the pilot sequences required to select the best beam
    pair is prohibitive when codebooks contain a large number of beams, as is
    the case in practice. To remedy this issue, some schemes exploit information
    about the user location to predict the best beam pair. However, these
    schemes (i) involve no measurements whatsoever, which generally results in a
    highly suboptimal predicted beam, and (ii) are not robust to localization
    errors. To address these limitations, this paper builds upon the  notion of
    radio map to develop two algorithms that attain a balance between the
    quality of the obtained beam pair and measurement overhead. The proposed
    algorithms predict the received power corresponding to each pair and
    measure just the $\numbeamstomeas$ pairs with highest prediction. While the
    first algorithm targets simplicity, the second one relies on a Bayesian
    approach to endow the prediction process with robustness to  localization
    error. The performance of both algorithms is shown to widely outperform
    existing methods using ray-tracing data.
\end{abstract}


\section{Introduction}
\label{sec:intro}

\begin{bullets}%
    \blt[Motivation]
    \begin{bullets}%
        \blt[6G] The demand for higher data rates, lower latency, and increased
        connectivity calls for the development of sixth-generation (6G)
        communication systems~\cite{letaief2019roadmap}. To support such high
        data rates, 6G systems are expected to heavily rely on the
        millimeter-wave (mmWave) frequency band, which spans from $30$ GHz to
        $300$ GHz.
        \blt[mmWave Challenges]The propagation impairments that characterize
        this band, such as high path loss,
        %
        \blt[Beamforming] render beamforming with large antenna arrays necessary
        at both the transmit and receive sides~\cite{wu2021environment}.
        %
        \blt[beam dicts]To reduce hardware costs, the beamforming vectors are
        chosen from a finite set, which is referred to as a beam codebook.
    \end{bullets}

    \blt[Literature]The problem of finding the right pair of transmit and
    receive beamforming vectors in the codebook is known as beam alignment.
    \begin{bullets}%
        \blt[Basic methods]
        \begin{bullets}%
            \blt[Exhaustive search]The simplest approach is the \emph{exhaustive
                search} method, where the gain of all possible pairs is measured in a
            \emph{measurement} (or training) \emph{phase} and the pair with greatest measured gain is selected
            for the \emph{data transmission phase}. Unfortunately, the duration of the
            measurement phase may be prohibitive since the number of pairs  is
            typically large in practice.
            \blt[Hierarchical search] To alleviate this limitation, \emph{hierarchical
                search} relies on codebooks of beams with nested radiation patterns
            to find the best pair using the bisection
            method~\cite{xiao2016hierarchical}. However, measurement noise may
            lead to erroneous decisions in early stages of this search, which
            can result in the selection of a highly suboptimal pair. This means
            that a long measurement  phase is required to
            sufficiently average out the noise and thereby obtain accurate gain estimates.
        \end{bullets}%

        \blt[Machine learning w/o location aware] Machine learning-based
        strategies have been proposed to reduce the measurement overhead of the
        aforementioned kinds of methods~\cite{cousik2021fast, ma2021deep,
            wang2022deep}.
        \begin{bullets}%
            \blt[Training based on RSS] For example,~\cite{cousik2021fast}
            predicts the optimal beam based on the measurements of a subset of
            beams. Inspired by hierarchical search,~\cite{wang2022deep} predicts
            the optimal narrow beam based on received signal strength (RSS)
            measurements of wide beams.
            \blt[Training based on CSI] In~\cite{ma2021deep},  the optimal beam
            is predicted based on the channel matrix measured in a
            lower-frequency band.
        \end{bullets}%

        \blt[Location aware]However, these methods do not take into account side
        information which can be useful to select the optimal pair with lower
        overhead.
        \begin{bullets}%
            \blt[methods]
            \begin{bullets}%
                \blt[Training based on side information]%
                To remedy this limitation,~\cite{zecchin2022lidar} predicts the optimal beam based on LiDAR measurements, the location of the
                base station (BS), and the location of the user equipment (UE).
                \blt[Maps] In~\cite{wu2021environment},  a radio map that provides the best beam pair index for
                every UE location is constructed.
            \end{bullets}%
            \blt[limitations]However, these methods suffer from two limitations:
            First, (L1) they provide a single beam pair, which is the one to be
            used for data transmission without previously measuring it. This means that
            prediction errors may result in a considerable performance
            degradation. Second, (L2) these methods are not robust to
            localization errors, which are significant in practice. \end{bullets}%
    \end{bullets}%


    \blt[contribution] The main contributions of this paper aim at  overcoming these limitations.
    \begin{bullets}%
        \blt[radio MAp BEam aLignment: MABEL] The first contribution is an  algorithm that  predicts the power of each beam pair by constructing a radio map for each of them. Predicting the power rather than the index, as in~\cite{wu2021environment}, results in a small collection of candidate beam pairs. Each beam pair is measured and the best one is used for transmission, which effectively solves (L1).
        \blt[LOcation-Robust bEam aligNment: LOREN] The second contribution is an algorithm that extends the first one to take into account localization errors. The pursued Bayesian approach results in a scheme that can cope not only with (L1) but also with (L2).
        \blt[Performance] The performance of both algorithms on  ray-tracing data is shown to be markedly superior to that of existing methods.
    \end{bullets}

    \blt[Organization] The rest of the paper is organized as follows. The
    system model and problem formulation are presented in
    Sec.~\ref{sec:mpf}. Strategies based on
    radio maps, including the first of the proposed algorithms, are discussed in Sec.~\ref{sec:rmba}. The second algorithm, which is robust to localization errors, is proposed in
    Sec.~\ref{sec:proposed_beamforming}. Finally,
    Secs.~\ref{sec:experiments} and~\ref{sec:conclusions} respectively provide simulation results and conclusions.
\end{bullets}

\section{Model and Problem Formulation}
\label{sec:mpf}

\begin{bullets}%
    \blt[model]
    \begin{bullets}%
        \blt[Area] Let $\region \in \rfield^\regiondim$ comprise the coordinates of all
        points in the spatial region of interest, where $\regiondim$ is typically 2 or 3.
        \blt[Network] Consider a mmWave system that comprises
        \blt[BS and UE]
        \begin{bullets}%
            \blt[BS] a BS  with $\numbsant$ antennas
            \blt[UE] a UE with  $\numueant$  antennas. The UE is   located at
            $\ueloc \in \region$.
        \end{bullets}
        \blt[Channel] Without loss of generality, the exposition focuses on the downlink, but the proposed methods carry over unaltered to the uplink. The channel matrix will be  denoted by
        $\channel \in \bbC^{\numueant \times \numbsant}$.

        \blt[Beams]
        \begin{bullets}
            \blt[Notation]The  beamforming vectors at the BS and UE are respectively denoted by $\bsbeam \in
                \bbC^{\numbsant \times 1}$,  and $\uebeam \in \bbC^{\numueant \times
                    1}$. Without loss of generality, it is assumed that both
            vectors are normalized, i.e., $\|\bsbeam\|^2 = \|\uebeam\|^2 = 1$.
            \blt[Finite number of beams]Vectors $\bsbeam$ and $\uebeam$ are respectively picked from the codebooks $\bsbeamset =
                \{\bsbeam_1, \ldots, \bsbeam_{\numbsbeam}\}$, and $\uebeamset =
                \{\uebeam_1, \ldots, \uebeam_{\numuebeam}\}$, where $\numbsbeam$ and
            $\numuebeam$ are respectively the numbers of transmit and receive beamforming
            vectors. Usually $\numbsbeam \leq \numbsant$ and
            $\numuebeam \leq \numueant$.
            \blt[Beamforming techniques] The proposed schemes are applicable to analog, digital, or hybrid  beamforming so long as the size of both codebooks is finite.
            \blt[Example of codebook] A typical example of a codebook is the \emph{discrete Fourier transform} (DFT) codebook, which for uniform
            linear arrays (ULAs) contains $\bsbeam_{\bsbeamind} =
                \frac{1}{\sqrt{\numbsant}}[1, e^{j\frac{2\pi}{\numbsant}\bsbeamind},
                    \ldots, e^{j\frac{2\pi}{\numbsant}\bsbeamind(\numbsant-1)}]\transpose$, $\bsbeamind=$ $1,\ldots,\numbsant$,
            and $\uebeam_{\uebeamind} = \frac{1}{\sqrt{\numueant}}[1,
                    e^{j\frac{2\pi}{\numueant}\uebeamind}, \ldots,
                    e^{j\frac{2\pi}{\numueant}\uebeamind(\numueant-1)}]\transpose$, $\uebeamind=$ $1,\ldots,\numueant$.
        \end{bullets}

        \blt[Received Signal] The  signal transmitted by the BS is denoted as $\sqrt{\bspower}\bssig \in \bbC^1$, where $\bspower$ is the transmit power and
        $\expectation[|\bssig|^2] = 1$.
        Then, the received signal at the
        UE  is given by
        \begin{align}
            \label{eq:received_signal}
            \uesig(\bsbeam, \uebeam) = \sqrt{\bspower}\uebeam^H\channel\bsbeam\bssig + \uebeam^H\uenoise,
        \end{align}
        where $\uenoise \in \bbC^{\numueant
                \times 1}$ is the additive noise at the receiver.
        \blt[Received power] Thus, one can define the RSS
        at the UE as $\uepower(\bsbeam, \uebeam) \define |\sqrt{\bspower}\uebeam^H\channel\bsbeam|^2$.
    \end{bullets}%

    \blt[Beam alignment problem]
    \begin{bullets}%
        \blt[Formulation] Given  $\bsbeamset$ and $\uebeamset$, the beam alignment
        problem is to find the beam pair that  maximizes the RSS:
        \begin{align}
            \label{eq:problem}
            \max_{\bsbeam \in \bsbeamset, \uebeam \in \uebeamset} \uepower(\bsbeam, \uebeam).
        \end{align}

        \blt[Approaches] As discussed in Sec.~\ref{sec:intro}, existing approaches use different kinds of information to  find the optimal pair.
        \blt[location-aware beam training] The present paper considers the problem of \emph{location-aware beam alignment}, in which
        \begin{bullets}%
            \blt[formulation]
            \begin{bullets}%
                \blt[UE location] a location estimate $\locest$ of the actual UE  location $\ueloc\in\region$ is given.
                \blt[data set] In addition, a set $\trainset$ of measurements of all beam pairs at $\trainlocnum$ locations is given, specifically
                $\trainset \define \{(\locest_{\trainlocind}, \measuepower_{\trainlocind, \bsbeamind, \uebeamind})\}_{\trainlocind, \bsbeamind, \uebeamind}$, where
                \begin{bullets}%
                    \blt $\measuepower_{\trainlocind, \bsbeamind, \uebeamind}$ is the measured RSS for the beam pair $(\bsbeam_{\bsbeamind}, \uebeam_{\uebeamind})$ at the $\trainlocind$-th location $\ueloc_{\trainlocind}\in \region$ and
                    \blt $\locest_{\trainlocind}$ is an estimate of $\ueloc_{\trainlocind}$, which in practice is obtained using localization techniques such as GPS.
                \end{bullets}%

            \end{bullets}%

            \blt[overview]The problem of location-aware beam alignment is
            addressed first in Sec.~\ref{sec:rmba} for the case where there is no
            location uncertainty, that is, $\locest = \ueloc$ and
            $\locest_\trainlocind = \ueloc_\trainlocind$ for all $\trainlocind$. The
            case where there is location uncertainty is addressed in
            Sec.~\ref{sec:proposed_beamforming}.
        \end{bullets}%
    \end{bullets}

\end{bullets}%

\section{Radio Map-based Beam Alignment}
\label{sec:rmba}

\begin{bullets}%

    \blt[overview]This section discusses approaches to solve the beam alignment problem using radio maps. To bypass the limitations of existing schemes, a new algorithm is proposed at the end of the section.

    \blt[LOS]
    \begin{bullets}%
        \blt[method]The simplest form of location-aware beam alignment algorithm arises when  propagation takes place in free space and the location and orientation of the BS and UE are known. In this case, one can compute the  angle of arrival (AoA) and the
        angle of departure (AoD) and choose the best beam pair based on the radiation pattern of each beam~\cite{maiberger2010location}.
        \blt[limitations \ra Error due to blockage]
        However, in practice, there are obstacles that can block the line of
        sight path, which renders this method ineffective.


    \end{bullets}%

    \blt[blockages \ra radio maps]
    \begin{bullets}%
        \blt[Overview] To address this limitation, radio maps can be used.
        \blt[intro to radio maps]
        \begin{bullets}%
            \blt[Radio map] Radio maps provide a radio frequency (RF) metric at
            each spatial location of a geographical
            area~\cite{romero2022cartography,
                zeng2024tutorial,
                romero2023theoretical,
                yilmaz2013radio,
                shrestha2024empiricaljpaper,
                alayafeki2008cartography,teganya2020autoencoders,
                shrestha2022surveying,
                romero2017spectrummaps,
                romero2018blind
            }.
            Possible RF metrics include the RSS,
            channel gain, power spectral density, and so on.
            \blt[Construction]
            \begin{bullets}%
                \blt[overview] Radio maps are commonly estimated by interpolating measurements acquired across the area
                of interest.

                \begin{journalonly}

                    \blt[approaches] Given the measurements at the measurement
                    locations, the radio map can be estimated using various
                    approaches, including kriging, Gaussian process regression,
                    and K-nearest neighbors (KNN) interpolation, ect.~\cite{romero2022cartography}.

                \end{journalonly}
            \end{bullets}%

        \end{bullets}%

        \blt[Channel path map \cite{wu2021environment}] One approach for
        location-aware beam alignment using radio maps can be found
        in~\cite{wu2021environment}. The scheme proposed therein constructs
        radio maps of   the AoA, AoD, and gain of each path from the BS to every
        possible UE location.
        \begin{bullets}%
            \blt[Channel model]Using the Saleh-Valenzuela model~\cite{saleh1987statistical},
            \blt[approach]
            $\channel$ is reconstructed at $\ueloc$ based on these quantities and then
            used to estimate the RSS for each beam pair.
            \begin{journalonly}
                between the BS and the UE is modeled as~\cite{saleh1987statistical}
                \begin{align}
                    \label{eq:channel}
                    \channel = \sqrt{\numbsant\numueant}\sum_{\pathind=1}^{\numpath}\pathgain_{\pathind}\uesteer(\uesteerang[\pathind])\bssteer(\bssteerang[\pathind])^H,
                \end{align}
                where
                \begin{bullets}%
                    \blt[Path] $\numpath$ is the number of paths,
                    \blt[Gain] $\pathgain_{\pathind}$ is the complex gain of the $\pathind$-th
                    path, and
                    \blt[Angle] $\uesteerang[\pathind]$ and $\bssteerang[\pathind]$ are the
                    angles of arrival and departure of the $\pathind$-th path, respectively.
                \end{bullets}
            \end{journalonly}
            \blt[limitations]
            \begin{bullets}%
                \blt[interp. path parameters] Unfortunately, spatially interpolating the aforementioned path parameters (AoAs, AoDs, and gains) to evaluate the radio maps is
                challenging because one cannot readily determine the correspondence between paths at different locations.

            \end{bullets}%

        \end{bullets}%

        \blt[Beam index map]
        \begin{bullets}%
            \blt[approach]To sidestep this challenge,  \cite{wu2021environment} also proposes
            constructing a ``beam index map" (BIM)  that provides the  index of the best beam pair
            for every UE location. This  beam  is directly utilized in the transmission phase without any measurement phase.
            \blt[limitations]
            \begin{bullets}%
                \blt[only one beam] Unfortunately, although this eliminates the measurement overhead,  the  beam provided by the map
                may be highly suboptimal, especially in areas where $\trainset$ contains few measurement locations.
                \blt[KNN] Also, this approach applies K-nearest neighbor (KNN) interpolation to the beam indices, which seems
                problematic in practice as beam indices are discrete and,
                therefore,  their averages may not capture the relations between the
                radiation patterns of the beams in the codebook.

            \end{bullets}%

        \end{bullets}%

        \blt[Radio map per beam pair]
        \begin{bullets}%
            \blt[motivation] To address these limitations, the first algorithm
            proposed here  constructs a radio map for each beam pair.
            %
            \blt[Problem]
            \begin{bullets}%
                \blt[given]Specifically, for each $(\bsbeamind,
                    \uebeamind)$,
                the data  $\{( \ueloc_{\trainlocind}, \measuepower_{\trainlocind,
                        \bsbeamind, \uebeamind})\}_{\trainlocind}$
                \blt[requested] is used to construct a radio map that provides an  estimate of the RSS $\uepower_{\bsbeamind, \uebeamind}$ at each possible UE location $\ueloc$.
            \end{bullets}%
            \blt[figure] Fig.~\ref{fig:radio_map} shows two examples of radio
            maps  for two different beam pairs. As can be seen, the RSS  is location-dependent and
            varies differently across space for  different beam pairs.
            \begin{figure}[t!]
                \centering
                \includegraphics[width=\columnwidth]{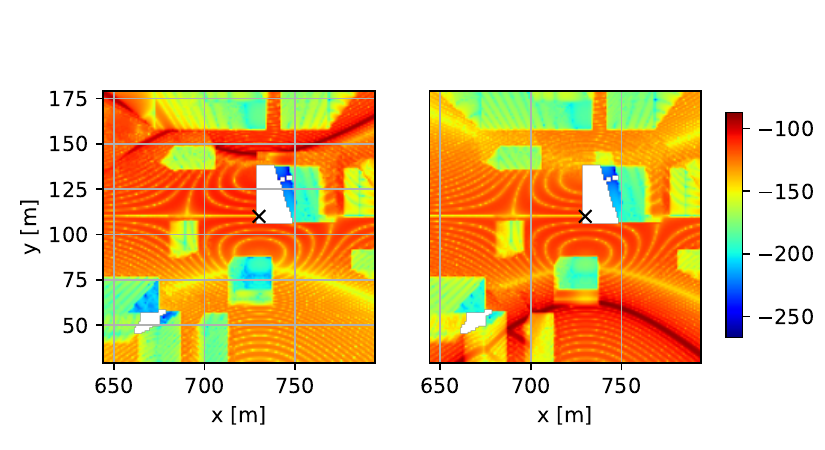}
                \caption{Examples of radio maps for two beam pairs. The
                    black crosses  represent the BS location. }
                \label{fig:radio_map}
            \end{figure}

            \blt[map construction] These $\numbsbeam\numuebeam$ radio maps can be constructed using any technique for power map estimation~\cite{romero2022cartography}. %
            \blt[memory complexity]
            However, since $\numbsbeam\numuebeam$  may be large, it is desirable to keep complexity low.
            To this end, KNN with 1 neighbor may be suitable since it just requires storing $\trainset$ and finding $\argmin_\trainlocind \|\ueloc - \ueloc_{\trainlocind}\|$ for each new $\ueloc$. This computation need not be performed for each beam pair but once for all pairs.
            %
            \blt[beam training] After  predicting the power of each beam pair at
            $\ueloc$ using the radio maps, the $\numbeamstomeas$ beam pairs with
            the highest power estimates are measured in the measurement phase. Based on
            the resulting measurements, the best of these  pairs is selected for the
            data transmission phase.

            \blt[name] This algorithm will be refered to as \emph{radio MAp BEam
                aLignment} (MABEL).

        \end{bullets}%

    \end{bullets}%

\end{bullets}%

\section{Radio Map-based Beam Alignment with Location Uncertainty}
\label{sec:proposed_beamforming}
\begin{figure}[t!]
    \centering
    \begin{tikzpicture}[->,>=stealth',shorten >=1pt,auto,node distance=1.1cm,
            thick, main node/.style={circle,draw,font=\sffamily\small, minimum size=0.8cm}]
        \node[main node] (1) {$\ueloc$};
        \node[main node] (2) [right of=1] {$\locest$};
        \node[main node] (3) [below of=1] {$\envir$};
        \node[main node] (4) [below of=2] {$\uepower$};
        \node[main node] (5) [below of=3] {$\ueloc_{\trainlocind}$};
        \node[main node] (6) [below of=4] {$\uepower_{\trainlocind}$};
        \node[main node] (7) [right of=6] {$\measuepower_{\trainlocind}$};
        \node[main node] (8) [below of=6] {$\locest_{\trainlocind}$};
        \node (9) [below right of=7] {$\trainlocind = 1, \ldots, \trainlocnum$};

        \path[every node/.style={font=\sffamily\small}]
        (1) edge node {} (2)
        (1) edge node {} (4)
        (3) edge node {} (4)
        (3) edge node {} (6)
        (5) edge node {} (6)
        (6) edge node {} (7)
        (5) edge node {} (8);
        \node [fit=(5)(6)(7)(8)(9), draw, rounded corners=6pt,rectangle] {};
    \end{tikzpicture}
    \caption{Graphical model of the system. \begin{journalonly}
            This model is used to predict
            the true map at locations for which the measurement is not available.
        \end{journalonly}}
    \label{fig:model}
\end{figure}

\begin{journalonly}
    \begin{figure}[t!]
        \begin{tikzpicture}[->,>=stealth',shorten >=1pt,auto,node distance=1.1cm,
                thick, main node/.style={circle,draw,font=\sffamily\small,
                        minimum size=0.8cm, align=center},
                text node/.style={font=\sffamily\small, minimum size=0.8cm,
                        align=center}]
            \node[main node] (1) {$\ueloc_{\trainlocind}$};
            \node[main node] (2) [right of=1] {$\locest_{\trainlocind}$};
            \node[main node] (3) [below of=1] {$\envir$};
            \node[main node] (4) [below of=2] {$\uepower_{\trainlocind}$};
            \node[main node] (5) [right of=4] {$\measuepower_{\trainlocind}$};
            \node[main node] (6) [below of=3] {$\ueloc_{\trainlocind'}$};
            \node[main node] (7) [below of=4] {$\uepower_{\trainlocind'}$};
            \node[main node] (8) [right of=7] {$\measuepower_{\trainlocind'}$};
            \node[main node] (9) [below of=7] {$\locest_{\trainlocind'}$};
            \node[text node] (10) [below right of=8] {$\trainlocind' = 1, \ldots, \trainlocnum$ \\ $\trainlocind'\neq \trainlocind$};

            \path[every node/.style={font=\sffamily\small}]
            (1) edge node {} (2)
            (1) edge node {} (4)
            (3) edge node {} (4)
            (4) edge node {} (5)
            (3) edge node {} (7)
            (6) edge node {} (7)
            (7) edge node {} (8)
            (6) edge node {} (9);
            \node [fit=(6)(7)(8)(9)(10), draw, rounded corners=6pt,rectangle] {};
        \end{tikzpicture}
        \caption{Special case of the model (just remove the test location). This
            model is used to predict the true map at the locations for which the
            measurement is available.}
        \label{fig:model_2}
    \end{figure}
\end{journalonly}

\begin{bullets}%
    \blt[overview]The approaches in Sec.~\ref{sec:rmba} assume that the UE
    location is perfectly known, but this is not the case in practice. Location error can be caused e.g. by multipath, the
    inaccuracy of the localization system, and the mobility of the UE. This section  builds upon  Bayesian principles to develop an improved version of MABEL where the interpolation technique used to construct the radio maps is robust to location errors.

    \blt[model] To this end, a probabilistic model must be adopted.
    \begin{bullets}%
        \blt[overview \ra graph] Such a model is represented by the Bayesian
        network in Fig.~\ref{fig:model}. Recall that Bayesian networks  constitute a
        special kind of graphical model~\cite[Ch. 8]{bishop2006}. This network
        captures the conditional independence relations among the considered random
        variables. The subscripts corresponding to the beam indices have been
        omitted to simplify notation. The procedure described in this section
        needs to be applied for each beam pair.  Circles denote random
        variables, and arrows denote the conditional dependencies. The model is
        fully defined by specifying the probability density functions (PDFs) of
        each random variable conditioned on its parents on the graph, namely
        $\pdf(\ueloc)$, $\pdf(\locest|\ueloc)$, $\pdf(\envir)$,
        $\pdf(\uepower|\ueloc,\envir)$, $\pdf(\ueloc_{\trainlocind})$,
        $\pdf(\uepower_\trainlocind|\ueloc,\envir)$,
        $\pdf(\locest_\trainlocind|\ueloc_{\trainlocind})$ and
        $\pdf(\measuepower_\trainlocind|\uepower_\trainlocind)$. The product of
        these PDFs is the joint PDF of all the random variables in the model.
        \blt[location] A non-informative prior is adopted for $\pdf(\ueloc)$ whereas  is $\pdf(\locest|\ueloc)$ is multivariate Gaussian  $ \mathcal{N}(\locest|\ueloc,
            \locerrorcov)$. The same applies to $\pdf(\ueloc_{\trainlocind})$ and $\pdf(\locest_\trainlocind|\ueloc_{\trainlocind})$.
        \blt[environment] Variable $\envir$ captures the propagation environment as well as all parameters and locations of the transmitters. This variable is used to express the RSS as a deterministic function of the location and the environment: $\uepower = \envirfun(\envir,\ueloc)$ and
        $\uepower_\trainlocind = \envirfun(\envir,\ueloc_\trainlocind)$, which in turn determine $\pdf(\uepower|\ueloc,\envir)$ and
        $\pdf(\uepower_\trainlocind|\ueloc_\trainlocind,\envir)$. In a simulation, $\envir$ could play the role of the 3D model of the environment and $\envirfun$ could be the ray-tracing software. The variable $\envir$ is not explicitly modeled, which means that $\pdf(\envir)$ and $\envirfun$ will be unknown.
        \blt[measurements] Finally, $\pdf(\measuepower_\trainlocind|\uepower_\trainlocind)$ can be assumed  to be e.g. $\mathcal{N}(\measuepower_\trainlocind|\uepower_\trainlocind,\measnoisevar)$, where $\measnoisevar$ is the variance of the additive measurement noise.

    \end{bullets}%

    \blt[Algorithm]
    \begin{bullets}%


        \blt[mmse estimate] Ideally, one would like to obtain the minimum mean square error (MMSE) estimator for the RSS.
        Applying a well-known result in estimation
        theory~\cite{kay2}, the MMSE estimator is given by $\expectation\left[\uepower|\locest,\trainset\right]$.
        \blt[pdf] To obtain this conditional expectation, observe that
        \begin{salign}[eq:pdf]
            \pdf&\left(\uepower|\locest,\trainset\right)
            = \int_{\ueloc\in\region} \pdf\left(\uepower,\ueloc|\locest,\trainset\right)d\ueloc
            \\&= \int_{\ueloc\in\region} \pdf\left(\uepower|\ueloc,\locest,\trainset\right)\pdf\left(\ueloc|\locest,\trainset\right)d\ueloc
            \\&=
            \int_{\ueloc\in\region} \pdf\left(\uepower|\ueloc,\trainset\right)\pdf\left(\ueloc|\locest\right)d\ueloc
            \label{eq:pdf2}
            \\&\approx
            \frac{\sum_{\trainlocind=1}^{\trainlocnum}\pdf\left(\ueloc=\locest_{\trainlocind}|\locest\right)\pdf\left(\uepower|\ueloc=\locest_{\trainlocind},\trainset\right)}{\sum_{\trainlocind=1}^{\trainlocnum}\pdf\left(\ueloc=\locest_{\trainlocind}|\locest\right)}.
            \label{eq:pdf3}
        \end{salign}
        \blt[Explain]Here,
        \begin{bullets}%
            \blt \eqref{eq:pdf2} follows from the fact that $\uepower$
            is conditionally independent of $\locest$ given $\ueloc$ and $\pdf\left(\ueloc|\locest\right)=\pdf\left(\ueloc|\locest,\trainset\right)$,   which follows from the D-separation rules~\cite[Sec. 8.2.2]{bishop2006},
            \blt and \eqref{eq:pdf3} is just a discrete approximation of the integral.
        \end{bullets}

        \blt[expectation]From \eqref{eq:pdf}, it follows that
        \begin{salign}[eq:expectation]
            &\expectation\left[\uepower|\locest,\trainset\right]
            \approx
            \frac{\sum_{\trainlocind=1}^{\trainlocnum}\pdf\left(\ueloc=\locest_{\trainlocind}|\locest\right)\expectation\left[\uepower|\ueloc=\locest_{\trainlocind},\trainset\right]}{\sum_{\trainlocind=1}^{\trainlocnum}\pdf\left(\ueloc=\locest_{\trainlocind}|\locest\right)}
            \\&=\frac{\sum_{\trainlocind=1}^{\trainlocnum}\exp{\left(-\frac{1}{2}(\locest_{\trainlocind}-\locest)\transpose\locerrorcov^{-1}(\locest_{\trainlocind}-\locest)\right)}\expectation\left[\uepower|\ueloc=\locest_{\trainlocind},\trainset\right]
            }{
                \sum_{\trainlocind=1}^{\trainlocnum}\exp{\left(-\frac{1}{2}(\locest_{\trainlocind}-\locest)\transpose\locerrorcov^{-1}(\locest_{\trainlocind}-\locest)\right)}},
        \end{salign}
        where it was used that
        \begin{salign}[eq:powestrhat]
            \pdf\left(\ueloc=\locest_{\trainlocind}|\locest\right)&= \frac{\pdf\left(\ueloc=\locest_{\trainlocind},\locest\right)}
            {\pdf\left(\locest\right)}
            = \frac{\pdf\left(\locest|\ueloc=\locest_{\trainlocind}\right)\pdf\left(\ueloc=\locest_{\trainlocind}\right)}
            {\pdf\left(\locest\right)}
            \\&=\pdf\left(\locest|\ueloc=\locest_{\trainlocind}\right)
        \end{salign}
        by assuming uniform priors for $\ueloc$ and $\locest$.
        \blt[radio map]Observe that \eqref{eq:powestrhat}, when seen as a function of $\locest$, is a radio map that provides the RSS of a beam pair.

        \blt[cases] The rest of this section approximates $\expectation\left[\uepower|\ueloc=\locest_{\trainlocind},\trainset\right]$ in the cases without and with localization error in the locations in $\trainset$.
        \begin{bullets}%
            \blt[Algorithm for $\ueloc_{\trainlocind} =
                    \locest_{\trainlocind}$]
            \begin{bullets}%
                \blt[overview] Suppose first that there is no localization error in these locations, i.e., $\ueloc_{\trainlocind} =
                    \locest_{\trainlocind}, \trainlocind = 1, \ldots, \trainlocnum$.
                \blt[cond. mean] As a result
                \begin{salign}[eq:trainexpectation]
                    & \expectation \left[\uepower|\ueloc =
                        \locest_{\trainlocind},\trainset\right] =
                    \expectation\left[\uepower|\ueloc=\ueloc_{\trainlocind},\trainset\right]
                    \\&\overset{A}{=}
                    \expectation\left[\uepower_{\trainlocind}|\ueloc=\ueloc_{\trainlocind},\trainset\right]
                    \overset{B}{=} \expectation\left[\uepower_{\trainlocind}|\trainset\right]
                    \\& =
                    \expectation\left[\uepower_{\trainlocind}|\ueloc_{\trainlocind},\measuepower_{\trainlocind},\trainset_{\trainlocind}\right]
                    \approx
                    \label{eq:trainexpectationapprox}
                    \expectation\left[\uepower_{\trainlocind}|\measuepower_{\trainlocind}\right],
                \end{salign}
                where
                \begin{bullets}%
                    \blt (A) follows from the fact that if  $\ueloc=\ueloc_\trainlocind$, then $\uepower = \envirfun(\envir,\ueloc)=\envirfun(\envir,\ueloc_\trainlocind)=\uepower_\trainlocind$,
                    \blt (B) follows from the D-separation rules,
                    \blt $\trainset_{\trainlocind} \define \trainset\setminus \{(\locest_{\trainlocind},
                        \measuepower_{\trainlocind})\}$,
                    \blt and the approximation in \eqref{eq:trainexpectationapprox} is explained in Appendix~\ref{app:exp}.
                \end{bullets}%
                \blt[example: Gaussian prior] If, for example,  $\uepower$ has PDF $\mathcal{N}(\uepower|\powermean,\powervar)$ and  is independent of $\measnoise$, then, it is easy to show that
                \begin{align}
                    \label{eq:trainexpectationapprox2bb}
                    \expectation\left[\uepower_{\trainlocind}|\measuepower_{\trainlocind}\right]
                    = \powermean + \frac{
                        \powervar
                    }{\powervar
                        + \measnoisevar
                    }(
                    \measuepower_{\trainlocind}
                    -
                    \powermean
                    ).
                \end{align}
            \end{bullets}
            \blt[Algorithm for $\bm r_l \neq \hbm r_l$]
            \begin{bullets}%
                \blt[Error in training locations] On the other hand, in the
                presence of localization error, one needs to introduce further
                approximations. Start by noting that
                \begin{salign}[eq:MMSEtrain]
                    &\expectation\left[\uepower|\ueloc
                        =\locest_{\trainlocind},\trainset\right]
                    \\&=
                    \int
                    \expectation\left[\uepower|\locest,\ueloc
                        =\locest_{\trainlocind},\trainset\right]
                    \pdf\left(\locest|\ueloc
                    =\locest_{\trainlocind},\trainset\right)
                    d \locest
                    \\&\approx\frac{\sum_{\trainlocind'=1}^{\trainlocnum}\pdf\left(\locest=\locest_{\trainlocind'}|\ueloc
                        =\locest_{\trainlocind}\right)
                        \expectation\left[\uepower|\locest=\locest_{\trainlocind'},\ueloc
                            =\locest_{\trainlocind},\trainset\right]
                    }{\sum_{\trainlocind'=1}^{\trainlocnum}\pdf\left(\locest=\locest_{\trainlocind'}|\ueloc
                        =\locest_{\trainlocind}\right)},
                \end{salign}
                where $\pdf\left(\locest|\ueloc
                    =\locest_{\trainlocind},\trainset\right)=\pdf\left(\locest|\ueloc
                    =\locest_{\trainlocind}\right)$ follows from the D-separation rules.

                To obtain $\expectation\left[\uepower|\locest=\locest_{\trainlocind'},\ueloc
                        =\locest_{\trainlocind},\trainset\right]$,
                suppose that, if $\locest=\locest_i$, then  $\ueloc = \ueloc_i$. This assumption is justified if (i) $\locest$ (resp. $\locest_i$) is a deterministic function of $\ueloc$ (resp. $\ueloc_i$) and the environment $\envir$, and (ii) this function is injective for each $\envir$. In practice, these two conditions may not exactly hold but they constitute a reasonable approximation. It follows that, if $\locest=\locest_i$, then                  $\uepower = \envirfun(\envir,\ueloc)=\envirfun(\envir,\ueloc_i)=\uepower_i$.
                Therefore,
                \begin{salign}
                    &\expectation\left[\uepower|\locest=\locest_{\trainlocind'},\ueloc=\locest_{\trainlocind},\trainset\right]
                    \\&= \expectation\left[\uepower_{\trainlocind'}|\locest=\locest_{\trainlocind'},\ueloc
                        =\locest_{\trainlocind},\ueloc_{\trainlocind'}
                        =\locest_{\trainlocind},\trainset\right]
                    \\
                    \label{eq:trainexpectationapprox2ff}
                    &= \expectation\left[\uepower_{\trainlocind'}|\ueloc_{\trainlocind'}
                        =\locest_{\trainlocind},\trainset\right]
                    \\&=\expectation\left[\uepower_{\trainlocind'}|\ueloc_{\trainlocind'}
                        =\locest_{\trainlocind},
                        \locest_{\trainlocind'},\measuepower_{\trainlocind'},\trainset_{\trainlocind'}\right]
                    \\
                    \label{eq:trainexpectationapprox2gg}
                    &=\expectation\left[\uepower_{\trainlocind'}|\ueloc_{\trainlocind'}
                        =\locest_{\trainlocind},\measuepower_{\trainlocind'},\trainset_{\trainlocind'}\right]
                    \\&
                    \label{eq:trainexpectationapprox2aa}
                    \approx \expectation\left[\uepower_{\trainlocind'}|\measuepower_{\trainlocind'}\right],
                    %
                \end{salign}
                where
                \begin{bullets}%
                    \blt \eqref{eq:trainexpectationapprox2ff} and \eqref{eq:trainexpectationapprox2gg} follow from the D-separation rules
                    \blt and \eqref{eq:trainexpectationapprox2aa} follows from the same arguments as in Appendix~\ref{app:exp}.
                \end{bullets}%
                Finally, $\expectation\left[\uepower_{\trainlocind'}|\measuepower_{\trainlocind'}\right]$ can be obtained e.g. as in \eqref{eq:trainexpectationapprox2bb}.
            \end{bullets}
        \end{bullets}%
    \end{bullets}%

    \blt[Testing time] Once the  RSS  of all  beam pairs has been estimated, the algorithm proceeds as MABEL: the $\numbeamstomeas$ beam pairs with the highest power estimates are measured in the measurement phase.
    \blt[name] The improved version of MABEL proposed in this section will be referred to as \emph{LOcation-Robust bEam aligNment} (LOREN).
\end{bullets}

\section{Simulation Results}
\label{sec:experiments}
\begin{bullets}%
    \blt[Overview]In this section, the performance of the proposed beam alignment strategies is
    evaluated through synthetic data from ray-tracing software and
    compared with existing beam alignment methods.

    \blt[Simulation setup]
    \begin{bullets}%
        \blt[Overview]To this end, a channel matrix is generated using
        ray-tracing for each UE location in a grid with 1 m spacing constructed
        in a square urban scenario with a side of $150$ m.
        \blt[frequency, bandwidth] The carrier frequency is $30$ GHz, the
        bandwidth is $1$ MHz,
        \blt[num antennas]
        $\numbsant=16$, and   $\numueant=4$.
        \blt[tx pow]The transmit power is $\bspower=0$ dBm, so that the values of RSS reported here can also be interpreted as gain.
        \blt[codebook] DFT codebooks are used in all cases.
        \blt[meas noise]To focus on spatial interpolation effects and the impact of the location error, no measurement noise is introduced.
        \blt[loc error]The location estimates are generated as
        $\locest \sim \mathcal{N}(\ueloc,
            \locnoisevar_\text{test}\eye_2)$ and
        $\locest_\trainlocind \sim \mathcal{N}(\ueloc_\trainlocind,
            \locnoisevar_\text{train}\eye_2)$.

    \end{bullets}

    \blt[Benchmarks]   The  following  benchmarks are considered:
    \begin{bullets}%
        \blt[Exhaustive search] (i) Exhaustive search beam alignment (ESBA): the best beam pair is chosen by measuring
        all possible beam pairs. Location information and $\trainset$ are not used.
        \blt[Hierarchical search] (ii) Hierarchical
        search beam alignment (HSBA) with
        $\log_2(\numbsant)$ levels at the BS and $\log_2(\numueant)$ levels at the UE. Each level involves measuring the four possible beam pairs in that level. Location information and $\trainset$ are not used.
        \blt[BIM] (iii) The BIM algorithm from~\cite{wu2021environment} with  $5$ neighbors. To improve its performance, instead of averaging beam indices, the optimal beams at the 5 nearest neighbors are measured.

    \end{bullets}

    \blt[evaluation]
    \begin{bullets}%
        \blt[procedure]
        \begin{bullets}%
            \blt[train/test locs]At every Monte Carlo iteration, $\trainlocnum$
            locations are randomly selected from the grid to form
            $\trainset$. A set of test locations $\ueloc$ is  randomly selected
            from the remaining grid points to average the performance metrics.
            \blt[beam index lists]For each of those testing locations, each
            algorithm returns a sorted list with the order in which the beams
            must be measured. ESBA and BIM must measure all the beam pairs in
            the list, MABEL and LOREN just measure  the first $\numbeamstomeas$
            pairs in the list to choose the best, and HSBA adaptively creates
            this list based on previous measurements.

        \end{bullets}%
        \blt[metrics]To analyze performance for all possible values of $\numbeamstomeas$ at the same time, the adopted metrics should therefore assess the ability
        of an algorithm to return strong beam pairs at the beginning of the
        list. In particular, the following metrics are considered:
        \begin{bullets}%
            \blt[best measured beam power so far] (i) \emph{The best measured
                beam power so far} (BMBPSF) is the sequence of maxima of the
            true (not measured) RSS of each beam in the order of the list
            returned by the tested algorithm. The faster it grows at
            the beginning, the better.
            \blt[Number of beam pairs] (ii) The number of measured beam pairs (NMBP) to
            reach an RSS that is within a given margin from the best beam pair.
        \end{bullets}
    \end{bullets}%

    \blt[Notation]In all experiments, LOREN assumes
    $\pdf(\locest|\ueloc)= \mathcal{N}(\locest|\ueloc,
        \bfstd_\text{test}^2\eye_2)$
    and
    $\pdf(\locest_\trainlocind|\ueloc_\trainlocind)= \mathcal{N}(\locest_\trainlocind|\ueloc_\trainlocind,
        \bfstd_\text{train}^2\eye_2)$, where  $\bfstd_\text{train}^2$
    and $\bfstd_\text{test}^2$
    need not equal the parameters $\locnoisevar_\text{train} $ and $\locnoisevar_\text{test}$, used to generate the data.

    \blt[Results]
    \begin{bullets}%
        \blt[Overview] The first step is to investigate the robustness of the considered algorithms with respect to the error in $\locest$. To this end, $\std_\text{train}$ will be initally set to $0$.
        \blt[exp.1-power vs num meas]
        \begin{bullets}%
            \blt[Results]
            Fig.~\ref{fig:exp1} shows the BMBPSF sequence for the considered algorithms when $\std_\text{test}=25$ m.
            \blt[Observations]
            \begin{bullets}
                \blt[1st beam pair]The left-most point in each curve provides
                the true RSS of the first measured pair.
                \begin{bullets}%
                    \blt[loc aware]
                    This RSS is greater for the algorithms based on radio maps
                    (BIM, MABEL, LOREN) than for the location-agnostic
                    algorithms (ESBA, HSBA), which shows the benefits of
                    exploiting location information.
                    \blt[ MABEL=BIM] The RSS of the first measured beam pair is
                    the same for BIM and MABEL. This is because the interpolation algorithm adopted by MABEL was
                    KNN with 1 neighbor and, therefore, the first
                    measured pair in both cases is the best beam pair at the nearest
                    neighbor in $\trainset$.
                \end{bullets}%
                \blt[last beam]
                \begin{bullets}%
                    \blt[Reach max power]
                    Except for BIM, which measures only 5 beams, the rest of
                    algorithms eventually reach the maximum RSS.
                    \blt[ESBA]ESBA is the slowest algorithm. This is expected as this is the
                    one that uses the smallest amount of information.
                    \blt[HSBA]HSBA reaches the maximum RSS faster than ESBA, which is reasonable since  it creates the list of beam pairs to measure  adaptively
                    based on previous measurements.
                    \blt[MABEL]The BMBPSF of MABEL grows faster than BIM, which
                    shows the benefits of interpolating RSS rather than just
                    mapping beam pair indices.
                    \blt[LOREN]The BMBPSF of LOREN  is the one that grows
                    fastest because of its robustness to the localization error
                    in $\locest$. Observe that it roughly reaches the maximum
                    power with just 10 measurements. This means that
                    $\numbeamstomeas$ can be set to 10. Although this number is
                    similar to the number of measurements required by HSBA to
                    reach the maximum in Fig.~\ref{fig:exp1}, the fact that  no
                    measurement noise is considered means that the curve for
                    HSBA is just an upper bound. In practice, the curve for HSBA
                    will be lower, meaning that it does not reach the maximum. In other words, when there is measurement noise,  the mean  power of the beam obtained by HSBA is lower than the mean  power of the best beam.
                \end{bullets}%
            \end{bullets}
            \begin{figure}
                \centering
                \includegraphics[width=\columnwidth]{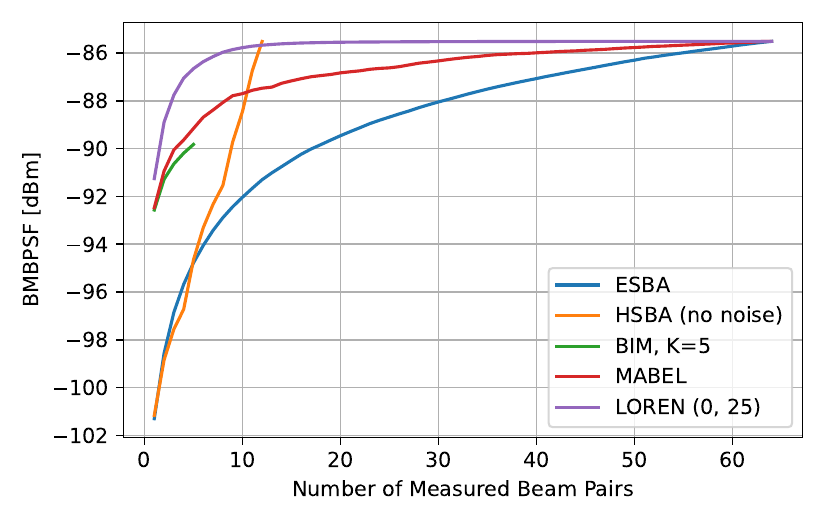}
                \caption{BMBPSF vs. the number of measurements when
                    $\std_\text{test}=25$ m and $\trainlocnum=500$. The legend entries of LOREN indicate
                    $(\bfstd_\text{train}, \bfstd_\text{test})$.}
                \label{fig:exp1}
            \end{figure}
        \end{bullets}

        \blt[exp.2-NMBP vs loc error]
        \begin{bullets}%
            \blt[motivation]Observe that LOREN requires knowledge of $\std_\text{train}$ and $\std_\text{test}$. In practice, these parameters may not be accurately known. Thus, it is important to assess the impact of a mismatch between the true and the assumed values of $\std_\text{train}$ and $\std_\text{test}$ on the performance of this algorithm.
            \blt[Results]To this end,
            Fig.~\ref{fig:exp2} plots
            NBPM with margin $1$ dB versus $\std_\text{test}$
            for several values of $\bfstd_\text{test}$.
            \blt[Observations]
            \begin{bullets}%
                \blt[all non-decreasing] As expected, the NMBP of the proposed algorithms increases with
                $\std_\text{test}$. The NMBP of ESBA and
                and HSBA is not affected by $\std_\text{test}$ because they do not utilize $\locest$.
                \blt[HSBA]Similarly to the previous experiment, the curve for HSBA is overly optimistic about the actual performance of this algorithm: in the presence of measurement noise, HSBA may not always find the best beam pair, which means that NMBP is undefined.
                \blt[LOREN]Observe that, for each $\std_\text{test}$, the smallest NMBP is obtained by LOREN when $\bfstd_\text{test}=\std_\text{test}$. Note also that the performance degradation is more significant if $\bfstd_\text{test}<\std_\text{test}$ than if $\bfstd_\text{test}>\std_\text{test}$, which motivates overestimating $\std_\text{test}$. For the specific setup in this experiment, it seems that $\bfstd_\text{test}=20$ m yields almost the best performance for all considered values of $\std_\text{test}$.
            \end{bullets}
            \begin{figure}
                \centering
                \includegraphics[width=\columnwidth]{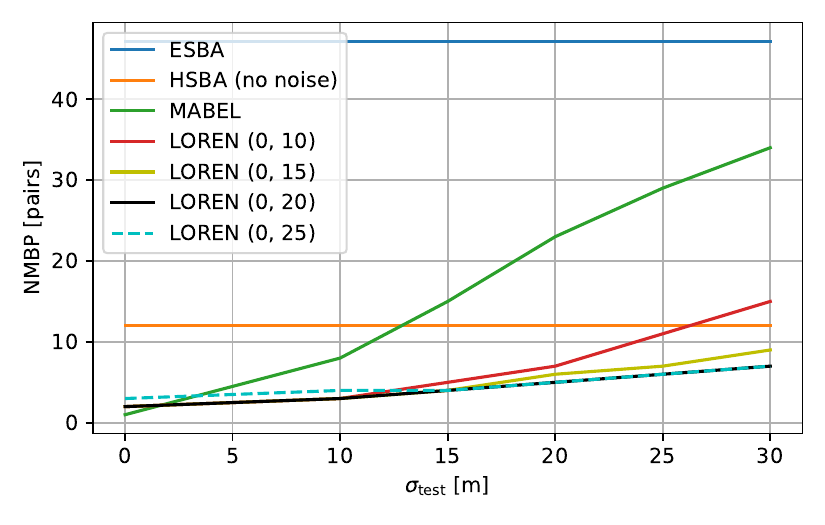}
                \caption{NMBP with margin $=1$ dB vs. $\std_\text{test}$. The
                    legend entries of LOREN indicate $(\bfstd_\text{train},
                        \bfstd_\text{test})$.}
                \label{fig:exp2}
            \end{figure}
        \end{bullets}

        \blt[exp.3-NMBP vs average minimum dist]
        \begin{bullets}%
            \blt[Motivation]Clearly, the performance of the algorithms using $\trainset$ is determined by $\trainlocnum$ or, equivalently, by the spatial density of the measurement locations in $\trainset$, which determines the cost of deploying these algorithms in practice.
            \blt[results]To investigate this effect, Fig.~\ref{fig:exp3} depicts the NMBP with margin $=1$ dB versus the average  distance between one  location in $\trainset$ and the nearest location in $\trainset$. Each point in the curves is obtained by setting a different $\trainlocnum$.
            \blt[Observations]
            \begin{bullets}%
                \blt[ESBA, HSBA] As before, NMBP for ESBA and HSBA remains constant because  these algorithms do not use $\trainset$. In the case of HSBA, this is again an overly optimistic curve shown here just as a reference.
                \blt[increases as min dist increases]As any algorithm based on radio maps, the performance of the proposed schemes improves with the spatial density of the  locations in $\trainset$.
                %
                \blt[10 m] Observe that with a spacing  of 10 m between locations in $\trainset$, LOREN and MABEL approximately achieve their optimal performance.
            \end{bullets}
            \begin{figure}
                \centering
                \includegraphics[width=\columnwidth]{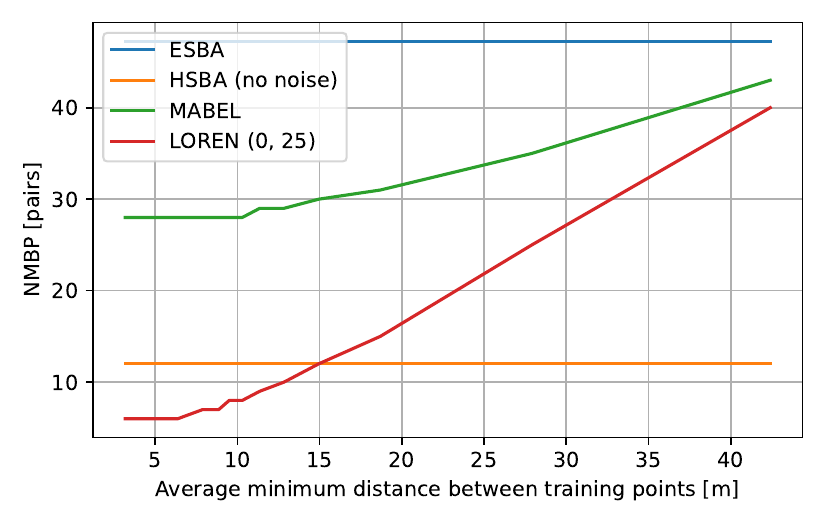}
                \caption{NMBP with a given margin $=1$ dB vs. the average minimum
                    distance between  locations in $\trainset$ when $\std_\text{test}=25$
                    m. The legend entries of LOREN indicate $(\bfstd_\text{train},
                        \bfstd_\text{test})$.}
                \label{fig:exp3}
            \end{figure}
        \end{bullets}

        \blt[exp.4-NMBP vs num bs ants]
        \begin{bullets}%
            \blt[Overview] To analyze how the  considered algorithms scale to systems with a large number  of  antennas, Fig.~\ref{fig:exp4} shows the NBPM with margin $=1$ dB vs. $\numbsant$.
            \blt[Observations]
            \begin{bullets}%
                \blt[all increase] As expected, the NBPM increases with the number
                of base station antennas  since the number of beam
                pairs increases.
                \blt[compare with benchmarks]LOREN results in the lowest  number of measurements to find a sufficiently good beam. Again, the curve for HSBA is just shown as a reference since it is obtained without measurement noise.
            \end{bullets}
            \begin{figure}
                \centering
                \includegraphics[width=\columnwidth]{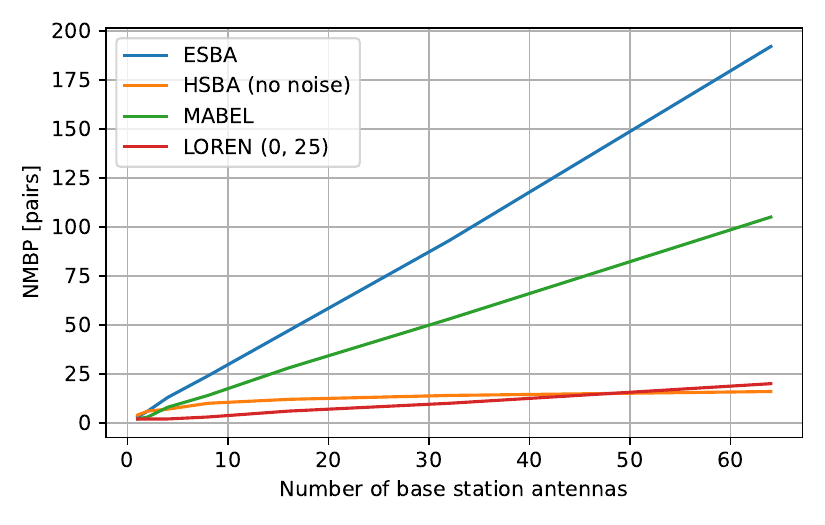}
                \caption{NMBP with a given margin $=1$ dB vs. number of base
                    station antennas when $\std_\text{test}=25$ m. The legend
                    entries of LOREN indicate $(\bfstd_\text{train},
                        \bfstd_\text{test})$.}
                \label{fig:exp4}
            \end{figure}
        \end{bullets}

        \blt[exp.5-power vs num meas, training locs error]
        \begin{bullets}%
            \blt[Overview] The last  experiment quantifies the effect of error in the
            locations in $\trainset$. To this end, the BMBPSF sequence is shown  for different $\std_\text{train}$ and $\std_\text{test}$ in
            Figs.~\ref{fig:exp5} and~\ref{fig:exp5_2}.
            \blt[Observations]
            \begin{bullets}%
                \blt[MABEL changes much]  It is observed
                that  the performance of
                MABEL is drastically deteriorated with both  $\std_\text{train}$ and $\std_\text{test}$.
                \blt[LOREN]In contrast,  LOREN is only slightly
                affected by the location error.
                \blt[parameters in LOREN]Besides, LOREN is not highly sensitive to the choice of  $\bfstd_\text{train}$ and $\bfstd_\text{test}$.
                \blt[parameter mismatch]For this algorithm, it is generally desirable to know the actual $\std_\text{train}$ and $\std_\text{test}$. Against intuition, Fig.~\ref{fig:exp5_2} shows that a mismatch in these parameters can sometimes lead to an insignificant performance improvement. The reason is that LOREN is not a true MMSE estimator but rather an approximation.

            \end{bullets}
            \begin{figure}
                \centering
                \includegraphics[width=\columnwidth]{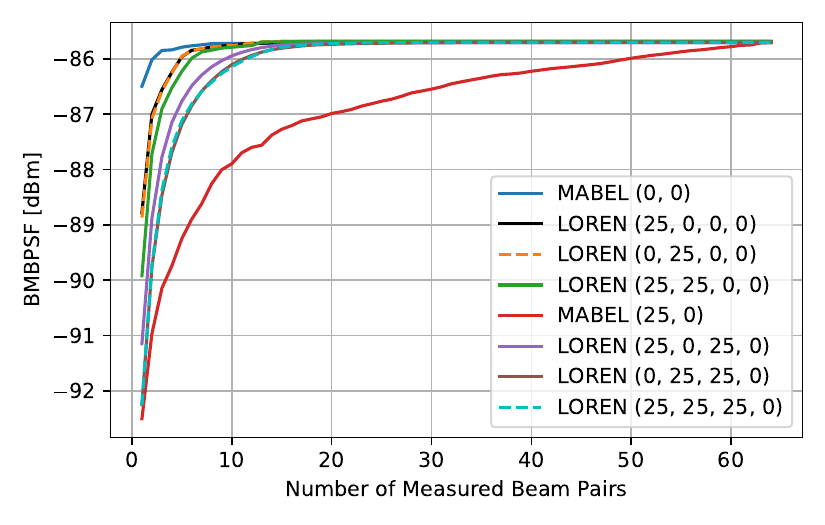}
                \caption{BMBPSF vs. the number of measured beam pairs for MABEL
                    ($\std_\text{train}, \std_\text{test}$) and LOREN
                    ($\bfstd_\text{train}, \bfstd_\text{test}, \std_\text{train},
                        \std_\text{test}$) when $\trainlocnum=500$.}
                \label{fig:exp5}
                \includegraphics[width=\columnwidth]{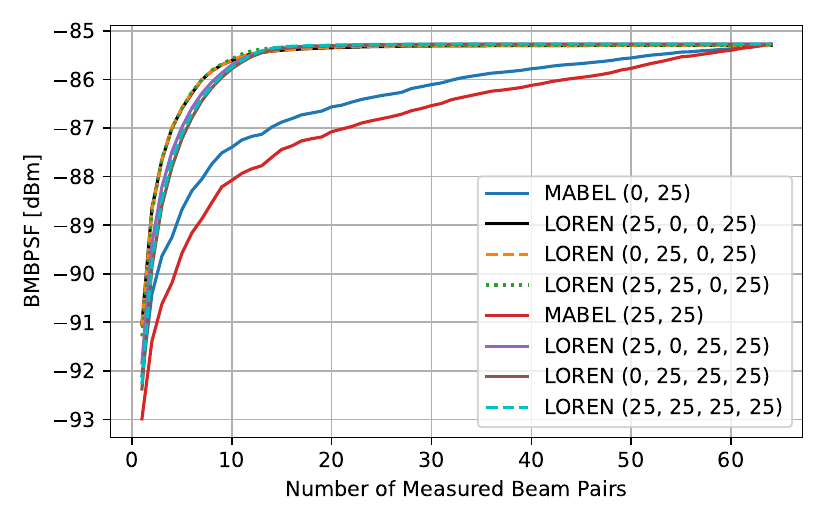}
                \caption{BMBPSF vs. the number of measured beam pairs for MABEL
                    ($\std_\text{train}, \std_\text{test}$) and LOREN
                    ($\bfstd_\text{train}, \bfstd_\text{test}, \std_\text{train},
                        \std_\text{test}$) when $\trainlocnum=500$.}
                \label{fig:exp5_2}
            \end{figure}
        \end{bullets}

    \end{bullets}
\end{bullets}

\section{Conclusions}
\label{sec:conclusions}

The problem of location-aware beam alignment was considered. To sidestep the limitations of existing algorithms when measurement locations are  exactly known, this paper proposed MABEL, which relies on radio maps that predict the RSS of each beam pair at the UE location. To accommodate location errors, LOREN extends MABEL by capitalizing on a Bayesian approach. Extensive simulations with ray-tracing data showcase the merits of these algorithms relative to the existing alternatives.

\appendices
\section{Approximation in \eqref{eq:trainexpectationapprox}}
\label{app:exp}

To approximate $\expectation\left[\uepower_{\trainlocind}|\ueloc_{\trainlocind},\measuepower_{\trainlocind},\trainset_{\trainlocind}\right]
    \approx
    \expectation\left[\uepower_{\trainlocind}|\measuepower_{\trainlocind}\right]$, start by noting that $
    \pdf(\uepower_{\trainlocind}, \ueloc_{\trainlocind}, \measuepower_{\trainlocind}, \trainset_{\trainlocind}) = \pdf(\uepower_{\trainlocind} | \ueloc_{\trainlocind}, \measuepower_{\trainlocind}, \trainset_{\trainlocind}) \pdf(\ueloc_{\trainlocind},
    \measuepower_{\trainlocind}, \trainset_{\trainlocind})$.
Since the $\envirfun$ is unknown,
it  will be assumed  that
$\pdf(\uepower_{\trainlocind} | \ueloc_{\trainlocind}, \measuepower_{\trainlocind}, \trainset_{\trainlocind}) \approx \pdf(\uepower_{\trainlocind} |
    \ueloc_{\trainlocind}, \measuepower_{\trainlocind})$,
which essentially means that the measurement $ \measuepower_{\trainlocind}$  at a certain location  $\ueloc_{\trainlocind}$ is much more informative about $\uepower_{\trainlocind} $ than the measurements at the remaining locations. The right-hand side of this expression can also be written as
\begin{salign}
    &\pdf(\uepower_{\trainlocind} | \ueloc_{\trainlocind}, \measuepower_{\trainlocind})
    \propto\int\pdf(\uepower_{\trainlocind}, \envir, \ueloc_{\trainlocind}, \measuepower_{\trainlocind})
    d\envir
    \\&=\int\pdf(\measuepower_{\trainlocind} | \uepower_{\trainlocind},  \envir, \ueloc_{\trainlocind})\pdf(\uepower_{\trainlocind} |
    \envir, \ueloc_{\trainlocind})\pdf(\envir)\pdf(\ueloc_{\trainlocind}) d\envir
    \\&=\int\pdf(\measuepower_{\trainlocind} | \uepower_{\trainlocind})\pdf(\uepower_{\trainlocind} |
    \envir, \ueloc_{\trainlocind})\pdf(\envir)\pdf(\ueloc_{\trainlocind}) d\envir
    \label{eq:approx1}
    \\&=\pdf(\measuepower_{\trainlocind} | \uepower_{\trainlocind})\pdf(\ueloc_{\trainlocind})\int \pdf(\uepower_{\trainlocind} |
    \envir, \ueloc_{\trainlocind})\pdf(\envir)d\envir
    \\&=\pdf(\ueloc_{\trainlocind})\pdf(\measuepower_{\trainlocind} |
    \uepower_{\trainlocind})\pdf(\uepower_{\trainlocind} |
    \ueloc_{\trainlocind}),
\end{salign}
where
\begin{bullets}%
    \blt[indep] \eqref{eq:approx1} follows from the D-separation rules.
\end{bullets}
Therefore,
\begin{salign}
    \pdf(\uepower_{\trainlocind}, \ueloc_{\trainlocind}, \measuepower_{\trainlocind}, \trainset_{\trainlocind})
    &\approx\frac{\pdf(\ueloc_{\trainlocind},
        \measuepower_{\trainlocind},\trainset_{\trainlocind})
        \pdf(\ueloc_{\trainlocind})
        \pdf(\measuepower_{\trainlocind} | \uepower_{\trainlocind})
        \pdf(\uepower_{\trainlocind} |
        \ueloc_{\trainlocind})}{ \int\pdf(\ueloc_{\trainlocind})\pdf(\measuepower_{\trainlocind} |
        \uepower_{\trainlocind})\pdf(\uepower_{\trainlocind} |
        \ueloc_{\trainlocind})  d\uepower_{\trainlocind}}
    \\&=\frac{\pdf(\ueloc_{\trainlocind},
        \measuepower_{\trainlocind},\trainset_{\trainlocind})\pdf(\measuepower_{\trainlocind} | \uepower_{\trainlocind})\pdf(\uepower_{\trainlocind} | \ueloc_{\trainlocind})}{
        \int\pdf(\measuepower_{\trainlocind} | \uepower_{\trainlocind})\pdf(\uepower_{\trainlocind} | \ueloc_{\trainlocind}) d\uepower_{\trainlocind}}.
\end{salign}

As a result,
\begin{salign}
    &\pdf(\uepower_{\trainlocind} | \ueloc_{\trainlocind},
    \measuepower_{\trainlocind}, \trainset_{\trainlocind}) =
    \frac{\pdf(\uepower_{\trainlocind}, \ueloc_{\trainlocind},
        \measuepower_{\trainlocind}, \trainset_{\trainlocind})}{\int
        \pdf(\uepower_{\trainlocind}, \ueloc_{\trainlocind},
        \measuepower_{\trainlocind}, \trainset_{\trainlocind})
        d\uepower_{\trainlocind}}
    \\&\approx \frac{\pdf(\measuepower_{\trainlocind} |
        \uepower_{\trainlocind})\pdf(\uepower_{\trainlocind} |
        \ueloc_{\trainlocind})}{\int \pdf(\measuepower_{\trainlocind} |
        \uepower_{\trainlocind})\pdf(\uepower_{\trainlocind} |
        \ueloc_{\trainlocind}) d\uepower_{\trainlocind}}
    \overset{A}{=
    } \frac{\pdf(\measuepower_{\trainlocind} |
        \uepower_{\trainlocind})\pdf(\uepower_{\trainlocind})}{\int
        \pdf(\measuepower_{\trainlocind} |
        \uepower_{\trainlocind})\pdf(\uepower_{\trainlocind})
        d\uepower_{\trainlocind}}
    \\&=\pdf(\uepower_{\trainlocind} | \measuepower_{\trainlocind}),
\end{salign}
where (A) follows by assuming that
$\pdf(\uepower_{\trainlocind} |
    \ueloc_{\trainlocind})=\pdf(\uepower_{\trainlocind})$, which means that the location is not informative about the RSS in the absence of further information. In other words, the medium is uniform and the set of possible environments $\envir$ is large enough not to favour any location over the others. This establishes the targeted approximation.
\begin{journalonly}
    \acom{explain in the main body?}
\end{journalonly}

\balance
\printmybibliography

\end{document}